\documentclass[prd,reprint,onecolumn,superscriptaddress,12pt]{revtex4-2}
\usepackage{anyfontsize}
\usepackage{graphicx}
\usepackage{amsfonts}
\usepackage{latexsym}
\usepackage{amsmath}
\usepackage[mathscr]{eucal}
\usepackage{palatino}
\usepackage{mathpazo}
\usepackage{textcomp}
\usepackage{booktabs}
\usepackage{dcolumn}
\usepackage{ragged2e}
\usepackage{hyperref}
\hypersetup{colorlinks,citecolor=blue}
\hypersetup{colorlinks=true,linkcolor=red,filecolor=magenta,urlcolor=blue}
\usepackage{amsmath}
\usepackage{xcolor}
\usepackage{orcidlink}
\usepackage[caption=false]{subfig}
\usepackage{commath}
\usepackage[italicdiff]{physics}

\newcommand{\ti}[1]{{\tilde{#1}}}
\begin{document}

\title{The extremal Reissner-Nordstr\"om black holes: an exact charged scalar quasiresonance.}

\author{David Senjaya} \email{davidsenjaya@protonmal.com} 
\affiliation{High Energy Physics Theory Group, Department of Physics, Faculty of Science, Chulalongkorn University, Bangkok 10330, Thailand}

\author{Supakchai Ponglertsakul}
\email{supakchai.p@gmail.com}
\affiliation{Strong Gravity Group, Department of Physics, Faculty of Science, Silpakorn University, Nakhon Pathom 73000, Thailand}

\date{\today}

\begin{abstract}
 In this letter, we present a novel exact scalar quasibound states solutions in the extremal Reissner-Norstr\"om black hole background. We start with the construction of the governing covariant relativistic scalar field equation, the Klein-Gordon equation in the extremal Reissner-Norstr\"om black hole background and applying the separation of variables anzat. The exact relativistic scalar wave's angular solution is found in terms of the spherical harmonics while the two independent radial wave solutions are, for the first time, exactly found and presented in terms of the double confluent Heun functions. The solutions are settled in the gravitational potential well and behave like an ingoing waves approaching black hole's horizon, vanishing when approaching infinity. The gravitationally bounded charged massive scalar fields are found to have quantized complex valued energy levels while imaginary energy levels are obtained for the charged massless scalar field, of both cases, indicating decaying states. Further investigation shows that the extreme Reissner-Nordst\"om black hole does not support scalar cloud. And with the help of the obtained exact radial solutions, the Hawking radiation of the extremal Reissner-Nordst\"om black hole is investigated and we find the zero temperature of the black hole's horizon.

\end{abstract}



\raggedbottom

\maketitle

\section{Introduction}

The efforts to formulate the theory of of quantum systems in a curved space-time background goes back to the year of 1920s-1930s. The study was carried out independently by Fock, Shcr\"odinger and Pauli to obtain the picture of the quantum mechanics in a
curved space \cite{Fock,Schrodinger,Pauli}. The effect of the space-time curvature on the Hydrogen atom has also been investigated in 1970s \cite{Audretsch}. It was found that the presence of the gravitational field represented by the space-time curvature alters the energy levels of the Hydrogen's electron. 

A concrete effort to investigate the analytical solutions of the relativistic scalar filed in a curved space-time by solving the Klein-Gordon equation for a massive field was carried out for the first time by Rowan and Stephenson in 1977 \cite{Rowan}. However, due to the complexity of the Klein-Gordon's radial equation, the analytical solutions were obtained only for asymptotic regions, i.e., close to the exterior of the event horizon and far away from the event horizon. The radial asymptotic solution found by Rowan and Stephenson is given in terms of the Hypergeometric functions that subjects to a polynomial condition, from where, the hydrogenic-like energy levels expression is obtained. Interestingly, scattering and bound states of scalar field on the Schwarzschild spacetime are obtained analytically in \cite{Li}. Quasibound states are computed explicitly in case of the Reissner-Nordstr\"om black holes \cite{Hod:2017gvn,Senjaya2,Abu-Saleem:2023ikx} and the Kerr black hole \cite{Hod:2015goa}. In addition, for acoustic black holes, the quasibound state solutions are obtained in many scenarios \cite{Vieira_2020,Vieira_2021,Vieira_2023}. Together with the advances in the research of the family of the Heun differential equations, we have developed a method to exactly solve the radial equation of the Klein-Gordon equation in several black hole backgrounds.  
Exact solutions of quasibound states of massless and massive neutral scalar field in various static spherically symmetric backgrounds have successfully been obtained in terms of either the confluent Heun or the general Heun functions \cite{Li,Vieira_2020,Vieira_2021,Vieira_2023,senjaya1,Senjaya2,senjaya5,senjaya4}.

On the otherhand, the failure of numerical methods to solve the Klein-Gordon equation with extremal black holes background is a well known problem \cite{Richartz,Joykutty}. The difficulty forces the numerical investigation stop short at the \textit{near-extremal} limit \cite{NE1,NE2,NE3,Cardoso:2003sw,Ponglertsakul:2018smo,Burikham:2020dfi,Wuthicharn:2019olp,Ponglertsakul:2020ufm}. Thus, in this paper, the charged scalar field in an extremal Reissner-Norstr\"om black hole background will be investigated analytically. 

We will start with the discussion of properties of the Reissner-Norstr\"om space-time in Sec~\ref{sec:bh} and the Klein-Gordon equation that is non-minimally coupled to the electromagnetic vector potential is constructed in Sec~\ref{sec:kg}. Using the ansatz of separation of variables, we solve the temporal and the angular parts respectively in terms of the harmonic function and the Spherical Harmonics. The radial part is then treated carefully and is brought to its normal form. We find that the radial equation of the charged scalar field in the extremal Reissner-Norstr\"om black hole background has exact solutions in terms of the double confluent Heun functions. By applying the polynomial conditions of the double confluent Heun function, the complex valued spectrum of the gravito-electromagnetically bound states are found for both massive and massless cases. This are discussed in Sec~\ref{subsec:energy}.

We will also examine the analytical energy levels expression in the so-called small black hole limit, given by the condition,
\begin{equation}
M_\mathrm{black\, hole}\ll\frac{m_\mathrm{Plank}^2 c^2}{E_{0}},    
\end{equation}
where $m_\mathrm{Plank}$ is Plank mass, $E_{0}$ is the scalar particle's rest energy and $c$ is the speed of light. In this particular condition, the imaginary part of the complex valued resonance frequencies are suppressed \cite{Anal4,Myung}, and we recover the expression found in \cite{Rowan} after setting $q=0$,
\begin{gather}
 \frac{E_n}{E_0}\approx 1-\frac{\mathcal{E}^2}{2n^2},\\
\mathcal{E}={\left(\frac{E_0 r_s}{\hbar c}\right)}^2.    
\end{gather}

Having the exact energy expression, we investigate the presence of the charged scalar cloud around the extremal Reissner-Norstr\"om black hole in Sec~\ref{sec:cloud} by imposing the $\tilde{\omega}=qQ$ and $\tilde{\omega}_I=0$ conditions and found out that the extremal Reissner-Norstr\"om black hole does not support scalar cloud.

In Sec~\ref{sec:hawk}, using the Damour-Ruffini method, we make use the obtained radial exact solutions to calculate the Hawking radiation of the Extremal Reissner-Norstr\"om black hole's horizon. We find out that the Extremal Reissner-Norstr\"om black hole's horizon does not radiate and it has zero Hawking temperature. We summarize our findings in Sec~\ref{sec:concl}.

\section{The Reissner-Nordstr\"om Black Hole}\label{sec:bh}
It did not take long after Einstein completed his general theory of relativity in 1915 for Karl Schwarzschild to find the first exact static spherically symmetric black hole solution \cite{Schw}. In the same year, Hans J. Reissner generalized the Schwarzschild's solution by including an electric charge into the black hole solution \cite{RN1}. Two years later, Gunnar Nordstr\"om independently obtained the same solution \cite{RN2}.

The Reissner-Nordstr\"om (RN) space-time describes the space-time geometry outside a static spherically symmetric charged massive body with mass $M$ and electric charge $Q$. The space-time is described by this following line element, 
\begin{gather}
ds^2=-f(r)dt^2+\frac{dr^2}{f(r)}+r^2d\theta^2
+r^2 \sin^2 \theta d\phi^2,\\ 
f(r)=\frac{(r-r_-)(r-r_+)}{r^2}, \label{metric}
\end{gather}
where the inner ($r_-$) and outer ($r_+$) horizons are given by
\begin{equation}
r_\pm = M \pm \sqrt{M^2-Q^2}. \label{rpm}
\end{equation}

Generally, the RN metric has two event horizons when $M > Q$. The RN metric develops to naked singularity as the event horizons become complex number i.e., $M < Q$. Interestingly, the event horizons are degenerate when gravitational mass $M$ equals electric charge $Q$ yielding the extremal black hole. Here in this article, we focus particularly on extremal RN black hole i.e., $r_-=r_+=M$. 

\section{The Klein-Gordon Equation}\label{sec:kg}
The quasibound states dynamics is governed by the covariant equation describing the behaviour of a relativistic charged scalar field in a curved spacetime is given by the Klein-Gordon equation,
\begin{equation}
D_{\nu}D^{\nu}\psi + \mu^2 \psi = 0, \nonumber
\end{equation}
where,
\begin{equation}
D_\nu=\nabla_{\nu} - iqA_\nu.
\end{equation}

Scalar field's mass and charge are denoted by $\mu$ and $q$ respectively. The electromagnetic potential is chosen to be,
\begin{align}
    A_{\nu} &= \left(\frac{Q}{r},0,0,0\right)
\end{align}
The scale field ansatz is given by,
\begin{align}
    \psi\left(t,r,\theta,\phi\right) &= e^{-i\omega t}R\left(r\right){Y_l^{m}}\left(\theta,\phi\right),
\end{align}
where $ Y_l^{m}\left(\theta,\phi\right)$ is the spherical harmonics.
Under extremal RN spacetime, the Klein-Gordon equation can be expressed as,
\begin{align}
    R'' + \frac{2}{r-r_+}R' + \left(\frac{r^2}{(r-r_+)^4}\left(\omega r + q Q\right)^2 - \frac{\mu^2r^2 + \ell(\ell+1)}{(r-r_+)^2}\right) R &= 0, \label{KGeq}
\end{align}
where $R'=dR/dr$ and $\ell(\ell+1)$ is the eigenvalue of spherical harmonic function. Now, we define a new variable,
\begin{align}
    x &\equiv \frac{r-r_+}{r_+}.
\end{align}
Thus, the event horizon is at $x=0$ and $x\to\infty$ as one approaches infinity. The Klien-Gordon equation becomes,
\begin{align}
 \frac{d^2R}{dx^2} + p\left(x\right)\frac{dR}{dx} +  
 +q\left(x\right)R=0, \label{radialeq}   
\end{align}
where,
\begin{align}
    p(x) &= \frac{2}{x}, \label{eqp} \\
    q(x) &= \left[\tilde{\omega}^2-\tilde{\mu}^2\right] + 2\left[\tilde{\omega}\left(2\tilde{\omega}+qQ\right)-\tilde{\mu}^2\right]\frac{1}{x} + \left[6\tilde{\omega}^2 + 6\tilde{\omega}qQ + q^2Q^2 \right. \nonumber \\
    & - \ell(\ell+1) - \tilde{\mu}^2 \left. \right]\frac{1}{x^2} + \left[2\left(2\tilde{\omega}+qQ\right)\left(\tilde{\omega}+qQ\right)\right]\frac{1}{x^3} + \left[\left(\tilde{\omega}+qQ\right)^2\right]\frac{1}{x^4}.
\end{align}

We define $\tilde{\omega}=\omega r_+$ and $\tilde{\mu}=\mu r_+$. Then, we transform the radial differential equation into the normal form by following the method described in Appendix~\ref{AppendixA}. The radial equation \eqref{radialeq} in the normal form reads,
\begin{align}
{\partial }^{{\rm 2}}_xY\left(x\right)+K\left(x\right)Y\left(x\right)&=0,    \label{radialnormalform1}
\end{align}
where $Y(x)\equiv x R(x)$. Since, $K(r)=-\frac{1}{2}\frac{dp}{dx}-\frac{1}{4}p^2+q$ (see Appendix~\ref{AppendixA}), equation \eqref{eqp} implies that $K(x)=q(x)$. This equation generally describes scalar perturbation on extremal RN background. In the following subsection, we shall impose specific boundary condition corresponding to quasibound states. 

\subsection{The Boundary Conditions}\label{subsec:bc}

The quasibound states are quasilocalized relativistic quantum states bound in the black hole's gravitation potential well. They have quasistationary energy levels and behave like an ingoing wave close to the black hole's horizon and vanishing at infinity. 

Let us consider the normal form of the radial equation \eqref{radialnormalform1}. Far away from the horizon, $x\to\infty$, the differential equation can be approximated in this following form,
\begin{equation}
    {\partial }^{{\rm 2}}_xY\left(x\right)+\left[\tilde{\omega}^2-\tilde{\mu}^2\right]Y\left(x\right)=0,   
\end{equation}
we obtain the decaying solution, we obtain a decaying solution,
\begin{equation}
  Y\left(x\right)=Ae^{-\sqrt{\tilde{\mu}^2-\tilde{\omega}^2}x},
\end{equation}
when $\ti{\omega} < \ti{\mu}$ is satisfied. 

Near the horizon, i.e. $x \to 0$, the radial equation \eqref{radialnormalform1} is then,
\begin{equation}
    {\partial }^{{\rm 2}}_xY\left(x\right)+\frac{\left(\tilde{\omega}+qQ\right)^2}{x^4}Y\left(x\right)=0,   
\end{equation}
where the solutions are obtained as follows,
\begin{align}
    Y\left(x\right)&= x\left[C_1 e^{i\left(\frac{\tilde{\omega}+qQ}{x}\right)}+C_2 e^{-i\left(\frac{\tilde{\omega}+qQ}{x}\right)}\right],
\end{align}
where $C_1,C_2$ are integration constants. Now, we define $k=\tilde{\omega}+qQ$ and redefining the radial variable $\tilde{x}=\frac{1}{x}$. We can set $C_1=0$ to obtain ingoing wave solution.


\subsection{The Radial Exact Solution}
By comparing \eqref{eqp} with the normal form of double confluent Heun differential equation \eqref{normalDCHeq} in Appendix~\ref{AppendixA}. One can identity the following,
\begin{align}
\epsilon &= 2\sqrt{\tilde{\mu}^2-\tilde{\omega}^2}, \\
\gamma &= 2i(\tilde{\omega}+qQ), \\
\delta &= 2 + 2iqQ + 4i\ti{\omega}, \label{delta} \\
\alpha &= 2\left(\frac{\epsilon}{2} -\ti{\mu}^2 + \left(2\ti{\omega} +qQ\right)\left(\ti{\omega}+\frac{i\epsilon}{2}\right)\right), \\
\beta &= \ti{\mu}^2 + \ell(\ell+1) - iqQ\left(1+\epsilon -2i\ti{\omega}\right) - 2\ti{\omega}\left(\ti{\omega} + i\left(1 + \frac{\epsilon}{2}\right)\right).
\end{align}

Here, we have successfully solved all of the double confluent Heun's parameters. Thus, the novel exact solutions of the Klein-Gordon equation in the extremal Reissner-Nordstr\"om background can be written as follows,
\begin{multline}
\psi=\psi_0 e^{-i\omega t}Y^{m_l}_l\left(\theta,\phi\right)\left[Ax^{\frac{1}{2}(\delta-2)}e^{\frac{1}{2}\left(\epsilon x-\frac{\gamma}{x}\right)}\operatorname{HeunD}\left(\beta,\alpha,\gamma ,\delta,\epsilon,x\right)\right. \\ \left.
+Bx^{-\frac{1}{2}(\delta-2)}e^{-\frac{1}{2}\left(\epsilon x-\frac{\gamma}{x}\right)}
\operatorname{HeunD}\left(-2+\beta+\delta,\alpha-2\epsilon,-\gamma ,4-\delta,-\epsilon,x\right)\right], \label{exactwavefunct}
\end{multline}
where HuenD is the double confluent Huen function. 

\subsection{The Scalar's Energy Quantization}\label{subsec:energy}
Now, let us apply the polynomial condition of the double confluent Heun as given in \eqref{HeunDPol}. The polynomialization of the radial solution is important since any infinite sequence of polynomials {$p_n$}, where $p_n$ having degree $n$ forms a basis for the infinite dimensional vector space of all polynomials. The set of the polynomial functions can be constructed as orthogonal basis by using the Gram-Schmidt orthogonalization. The polynomial condition terminates the infinite series of the radial solution at $n_r^{th}$ order and leads to quantization of relativistic scalar's energy as follows, 
\begin{equation}
 \frac{4\tilde{\omega}^2-2\tilde{\mu}^2+2\tilde{\omega} qQ}{2\sqrt{\tilde{\mu}^2-\tilde{\omega}^2}}+\left(1-\frac{4\tilde{\omega}^2+2q^2Q^2+6\tilde{\omega} qQ}{2i(\tilde{\omega}+qQ)}\right)=-n_r, \label{energylevelsexact}   
\end{equation}
where $n_r$ is the principal quantum number that can be connected with the angular momentum quantum number $\ell$ and the excitation number $N$ as $n_r=N+\ell$.

In addition, we investigate the energy quantization in the small black hole limit. Taking $\tilde{\omega}=\omega r_+\ll 1$, $qQ \ll 1$ and $\frac{\tilde{\omega}}{\tilde{\mu}} \approx 1$, for massive charged scalar field, a Hydrogenic energy levels expression is obtained, 
\begin{gather}
\tilde{\omega} {\rm =}{\tilde{\mu} }\sqrt{1-\frac{(\tilde{\mu} + qQ)^2}{{\left(n_r+1\right)}^2}}, \label{energylevelsapprox}
\end{gather}.

In the limit where $\frac{\left(\ti{\mu}+qQ\right)^2}{\left(n_r+1\right)^2}\ll 1$, one can expand \eqref{energylevelsapprox} as a Taylor Series as follows, 
\begin{equation}
\tilde{\omega}=\tilde{\mu}\left[1-\frac{(\tilde{\mu} + qQ)^2}{2{\left(n_r+1\right)}^2}-\frac{(\tilde{\mu} + qQ)^4}{8{\left(n_r+1\right)}^4}-\frac{(\tilde{\mu} + qQ)^6}{16{\left(n_r+1\right)}^6}+\dots \right].
\end{equation}

If $q$ is set to be zero, we recover the well-known energy levels of the so called gravitational atom,
\begin{gather}
\frac{\omega_n}{\mu}\approx 1-\frac{\kappa^2}{2(n_r+1)^2},   \kappa={\mu M}^2,
\end{gather}
that can be found in \cite{Daniel,Anal1,Anal2,Anal3,Anal10}. Analogous to the electronic transition that emits photon, graviton is emitted as level transition occurs in a gravitational atom \cite{Hu,Yang}. 

In the case of massless field $(\ti{\mu}=0)$, energy level can be solved as follows,
\begin{gather}
\tilde{\omega}= - \frac{qQ}{2} + \frac{\left(n_r+1\right)}{4}i. \label{masslessenergy}
\end{gather}

Clearly, the interaction between black hole and field dictates the real part of quasibound state frequency while the imaginary part solely depends on the principal quantum number $n_r$. The positive real part requires that $qQ < 0$, therefore, there is no superradiance in this case \cite{Benone:2015bst}. In addition, it is shown in \cite{HOD2012505} that extremal Reissner-Nordstr\"om black hole does not suffer from superradiant instability. In the absence of scalar field charge ($q=0$), this formula agrees with those obtained in \cite{Senjaya2}.
\begin{figure}[ht]
    \centering
    \includegraphics[scale=.8]{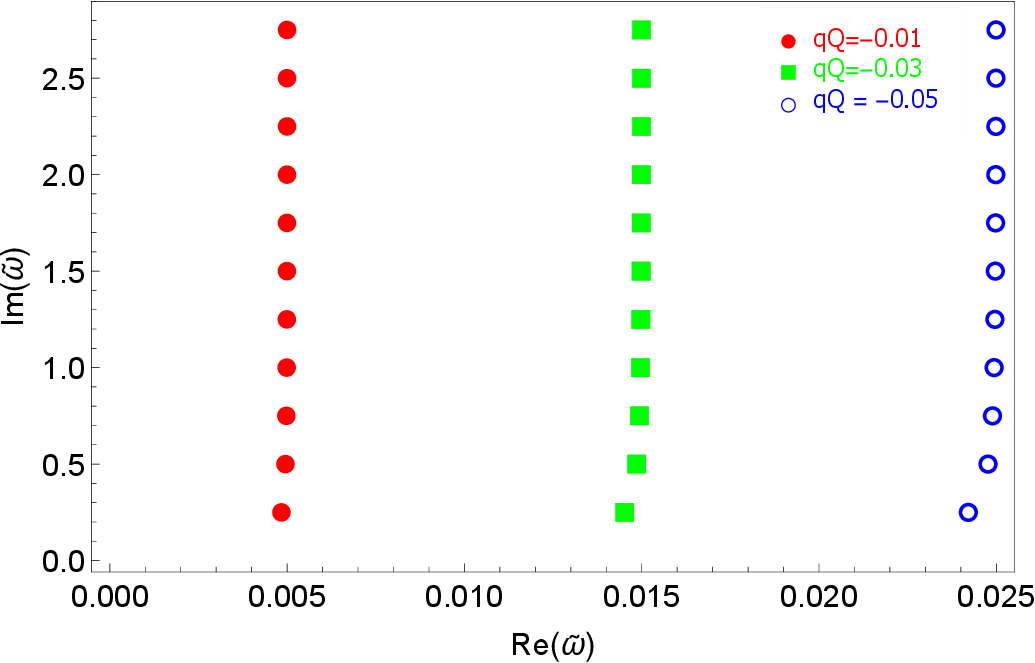} 
    \caption{Quasibound state's frequencies with $\ti{\mu}=0.1$ where $n_r=0-10$ increasing from bottom to top. } 
  \label{fig:qbs1}
\end{figure}

\begin{figure}[ht]
    \centering
    \includegraphics[scale=.45]{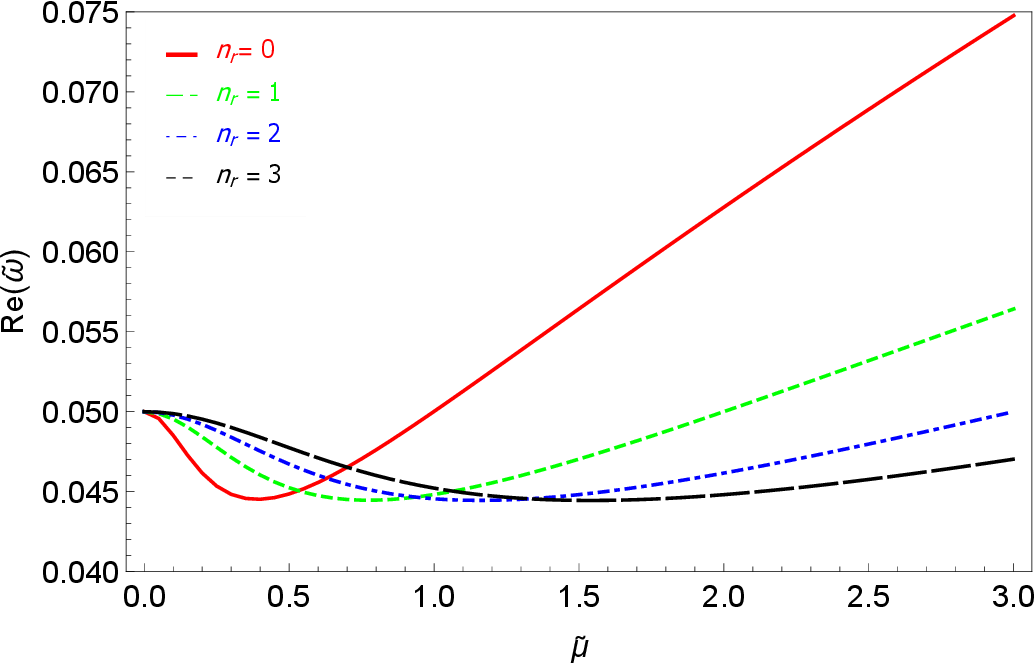}
    \includegraphics[scale=.45]{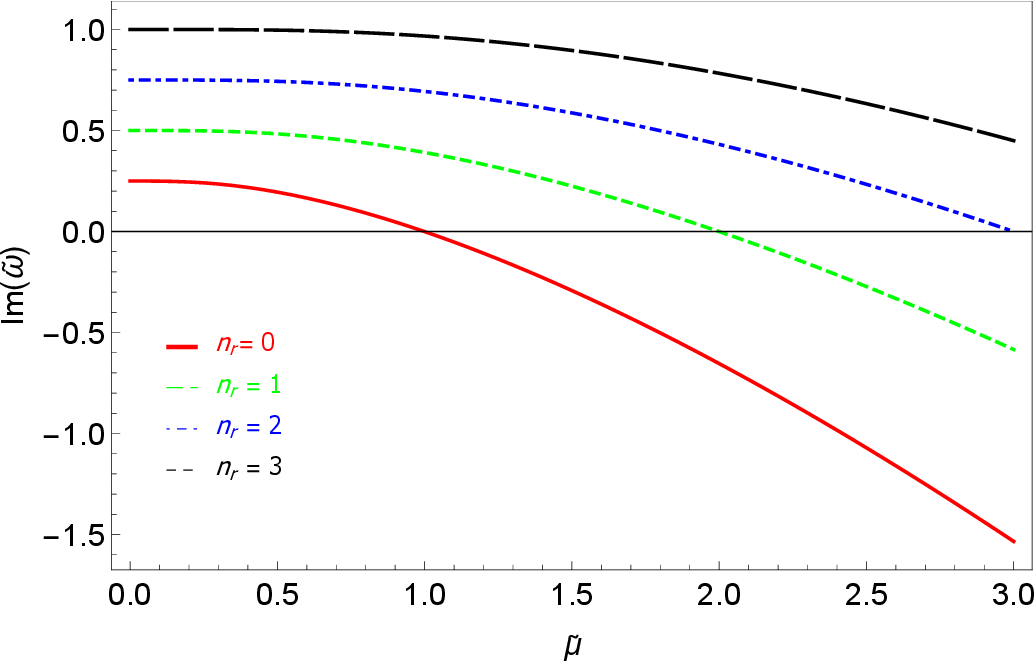}
    \caption{Quasibound state's frequencies as a function of $\ti{\mu}$ for fixed $qQ=-0.1$.} 
  \label{fig:qbs2}
\end{figure}

\begin{figure}[ht]
    \centering
    \includegraphics[scale=.45]{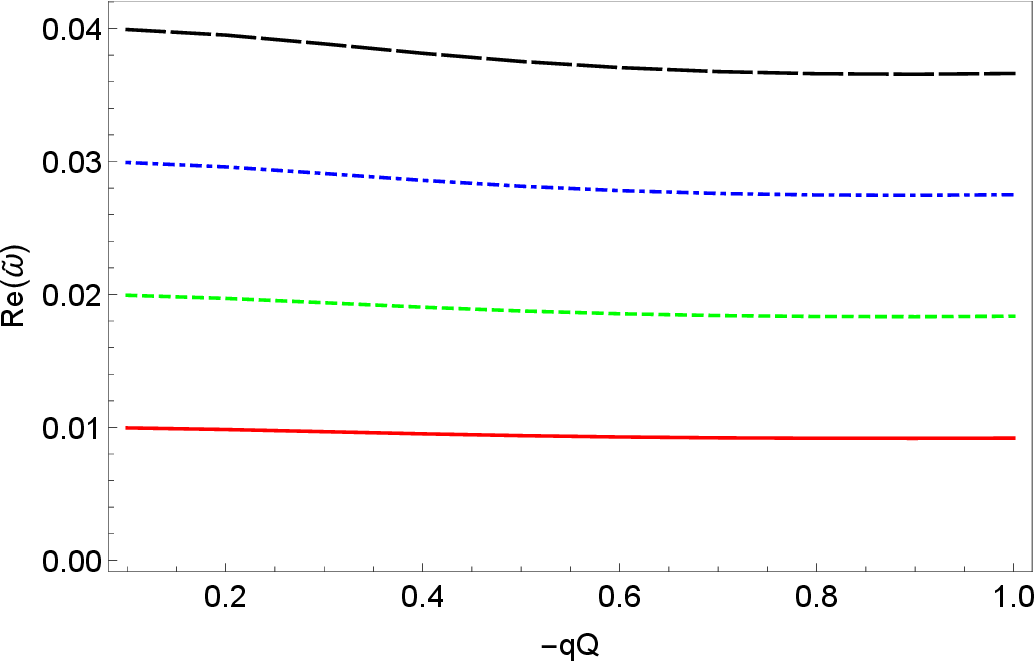}
    \includegraphics[scale=.45]{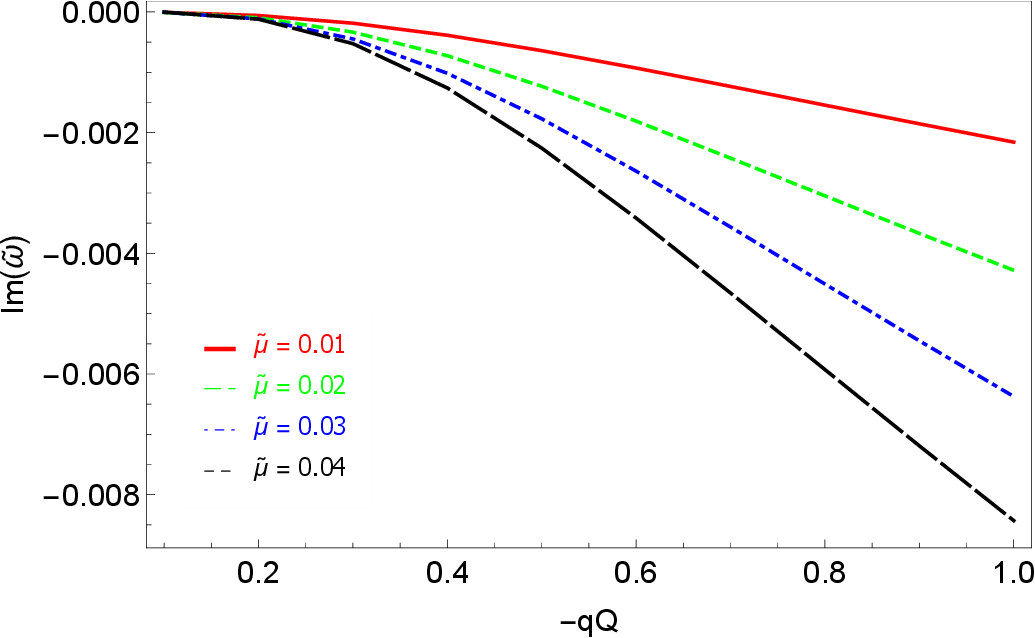}
    \caption{Quasibound state's frequencies as a function of $-qQ$ with $n_r=0$. The colored curves are denoted by the legend in the right plot.} 
  \label{fig:qbs3}
\end{figure}


In Fig.~\ref{fig:qbs1}, we plot quantized quasibound states frequencies in a complex $\Tilde{\omega}$ plane for a scalar field with mass $\Tilde{\mu}=0.1$ in the extremal Reissner-Nordstr\"om black hole background where the charge configurations are varied as $qQ=-0.01,-0.03,-0.05$. We present the quasibound states frequencies for $n_r=0,1,...,10$ where the lowest $n_r$ is always associated with lowest value of $Im(\tilde{\omega})$. The leap of the imaginary part of the quasibound states frequencies, $\Delta Im(\tilde{\omega})$, seems steady while the real part of the quasibound states frequencies significantly increase for lower $n_r$ and slightly increase for higher $n_r$, similar to the hydrogenic atom's electronic level transitions. 

In Fig.~\ref{fig:qbs2}, we separately present the real and imaginary quasibound states frequencies for a fixed charge configuration $qQ=-0.1$. We solve the $\Tilde{\omega}$ for various value of $\Tilde{\mu}$ with $n_r=0,1,2,3$. It is important to mention that the imaginary part of the quasibound states frequencies runs from positive, zero, and finally entering negative zone. The zero imaginary part of the frequencies indicates the existence of a scalar cloud which will be elaborated more in the following section.

In Fig.~\ref{fig:qbs3}, we present visualizations of the real and imaginary part of the quasibound states frequencies for various $qQ-\tilde{\mu}$ configuration. The $Re(\tilde{\omega})~\text{vs}~-qQ$ graph shows us that larger value of $-qQ$ causes larger drop of $Re(\tilde{\omega})$. Classically, this is understood as the adding up of the attractive electrostatic potential to the gravitational potential enhancing the binding energy of the scalar fields to the central black hole. The $Im(\tilde{\omega})~\text{vs}~-qQ$ graph also shows us that the quasibound states are stable since all values of the $Im(\tilde{\omega})$ are negative indicating decaying states. Moreover, we also notice that the value of $Im(\tilde{\omega})$ increases in magnitude following the increase in magnitude of $qQ$. This indicates that the power of the extremal charged black hole in absorbing scalar particles is increased with the help of the attractive electrostatic interaction. Also notice that for each value of $qQ$, larger $\Tilde{\mu}$ leads to more rapid decay.



\section{The Charged Scalar Cloud}\label{sec:cloud}
In this section, we will investigate the existence of the scalar cloud around the extremal Reissner-Nordstr\"om black hole. The scalar cloud is an stable configurations of a charged scalar field stationarily bound in a black hole background. The scalar cloud have a real valued frequency that matches the $qQ$ configuration as follows \cite{Garc,Hod3,Herdeiro:2014goa},
\begin{gather}
\tilde{\omega}=qQ,\\
\tilde{\omega}_I=0. \label{cond1}
\end{gather}

The first scalar cloud condition, classically, is understood as the equilibrium between gravitational-electrostatic force balance. The second energy condition indicates that there is no ingoing scalar flux impenetrates the outer horizon. Therefore, they do not grow or decay over time.

Applying the first condition to the exact energy expression \eqref{energylevelsexact}, we obtain,
\begin{equation}
 \frac{6\tilde{\omega}^2-2\tilde{\mu}^2}{2\sqrt{\tilde{\mu}^2-\tilde{\omega}^2}}+\left(1+3i\tilde{\omega}\right)=-n_r. \label{Cloud}  
\end{equation}
where the analytical solution must be complex valued, thus, violates the second condition of the scalar cloud. Now, let us check whether there exists a massless charged scalar cloud around the extremal Reissner-Nordstr\"om black hole by setting $\tilde{\mu}=0$ in the equation \eqref{Cloud}, here we obtain,
\begin{equation}
\tilde{\omega}=i\frac{(n_r+1)}{6}.
\end{equation}

Again, the solution is purely imaginary, violating the second criterion of the scalar cloud. Therefore, there is no solution for $\tilde{\mu}$ that satisfies the energy condition \eqref{cond1}. Thus, the charge scalar cloud surrounding an extreme Reissner-Nordstr\"om black hole does not exist. The same conclusion is obtained via asymptotical method and can be found in \cite{Garc}. 

However, it is possible to find a condition for a long-lived massive scalar cloud by taking $\tilde{\omega}=\omega r_+\ll 1$ and $\frac{\tilde{\omega}}{\tilde{\mu}} \approx 1$ limit of the \eqref{Cloud} and obtain a real valued energy levels expression as follows,
\begin{equation}
\tilde{\omega} = \tilde{\mu} \sqrt{1-\frac{4\tilde{\mu}^2}{(n_r+1)^2}}. \label{mindec}   
\end{equation}

Despite the extremal Reissner-Nordstr\"om black hole does not support scalar cloud. But, it is possible to find scalar configuration that minimizes the decay. 

\section{The Hawking Radiation}\label{sec:hawk}
In this section, the Hawking radiation coming out of the extremal Reissner-Nordstr\"om black hole's horizon will be investigated. With the exact solution of the radial wave function in hand, we can apply the Damour-Ruffini method \cite{Damour,Sannan} in order to derive the radiation distribution function without having to go through the black hole's thermodynamics. Let us start with rewriting the complete radial solution of the Klein-Gordon equation as follows,
\begin{multline}
R=Ax^{\frac{1}{2}(\delta-2)}e^{\frac{1}{2}\left(\epsilon x-\frac{\gamma}{x}\right)}\operatorname{HeunD}\left(\beta,\alpha,\gamma ,\delta,\epsilon,x\right)\\
+Bx^{-\frac{1}{2}(\delta-2)}e^{-\frac{1}{2}\left(\epsilon x-\frac{\gamma}{x}\right)}
\operatorname{HeunD}\left(-2+\beta+\delta,\alpha-2\epsilon,-\gamma ,4-\delta,-\epsilon,x\right).
\end{multline}

Approaching the horizon $r_+$ i.e.,$x\to 0$, the wave function can be expanded up to the first order of $x$, where, 
\begin{align}
e^{\frac{1}{2}\left(-\frac{\gamma}{x}\right)}\operatorname{HeunD}\left(\beta,\alpha,\gamma ,\delta,\epsilon,x\to0\right)  &\equiv \phi_1,\\
e^{\frac{1}{2}\left(\frac{\gamma}{x}\right)}
\operatorname{HeunD}\left(-2+\beta+\delta,\alpha-2\epsilon,-\gamma ,4-\delta,-\epsilon,x\to0\right) &\equiv \phi_2.   
\end{align} 

Notice that, the $\operatorname{HeunD}(x\to 0)\to\infty$ regardless the value of the parameter. Therefore, $\phi_1$ can be considerably finite while $\phi_2$ diverges as one moves closer to the horizon. Therefore, in the near horizon limit, the radial wave can be rewritten in this following form,
\begin{equation}
R=A\phi_1x^{\frac{1}{2}(\delta-2)}
+B\phi_2x^{-\frac{1}{2}(\delta-2)}.        
\end{equation}

The scalar's radial wave consists of two parts, i.e. $\psi_{+in}$ is the ingoing wave and $\psi_{+out}$ is the outgoing wave as follows, 
\begin{gather}
R =\left\{ \begin{array}{cc}
\psi_{+in}=B\phi_2x^{-\frac{1}{2}(\delta-2)}\\ 
\psi_{+out}= A\phi_1x^{\frac{1}{2}(\delta-2)}\end{array}
\right.,
\end{gather}
where after defining $\kappa=\frac{1}{2}(\delta-2)$, the combination of the radial and temporal solution leads to this following simple wave equation,
\begin{gather}
R =\left\{ \begin{array}{cc}
\psi_{+in}=B\phi_2e^{-i\left(\omega t+\kappa\ln{x}\right)}\\ 
\psi_{+out}= A\phi_1e^{-i\left(\omega t-\kappa \ln{x}\right)} \end{array}
\right.,
\end{gather}

Furthermore, we can investigate the ratio $\frac{\phi_2}{\phi_1}$ as follows,
\begin{align}
    \frac{\phi_2}{\phi_1}=\frac{\operatorname{HeunD}\left(-2+\beta+\delta,\alpha-2\epsilon,-\gamma ,4-\delta,-\epsilon,x\to0\right)}{\operatorname{HeunD}\left(\beta,\alpha,\gamma ,\delta,\epsilon,x\to0\right)} e^{\left(\frac{\gamma}{x}\right)},
\end{align}
and as the $\operatorname{HeunD}$ are series solutions, the behaviour of the quantity $ \frac{\phi_2}{\phi_1}$ is mostly determind by the exponential term. So, in the limit $x\to 0$, we obtain,
\begin{equation}
\lim_{x\to 0} \frac{\phi_2}{\phi_1} = \infty.
\end{equation}

Now, let us apply the Damour-Ruffini method as the following. Suppose we have an ingoing wave hitting the apparent horizon at $r_+$ and inducing a particle-antiparticle pair where the particle will enhance the reflected wave and the antiparticle will become the transmitted wave. Analytical continuation of the wave function $\psi\left(x\right)$ describes unique antiparticle state for a particular particle state and it is defined as $\psi_{-out}\equiv \psi_{+out}\left(x\to xe^{-i\pi}\right)$ is obtained simply by changing $x\to -x=xe^{-i\pi}$ as follows,
\begin{equation}
\begin{split}
\psi_{-out}&=A\phi_1\left(xe^{-i\pi}\right)^{\frac{1}{2}(\delta-2)},\\
&=\psi_{+out}e^{-\frac{1}{2}i\pi(\delta-2)}.
\end{split}
\end{equation}

We can therefore, consider the following ratio,
\begin{align}
{\left\lvert \frac{\psi_{-out}}{\psi_{+in}}\right\rvert }^2={\left\lvert \frac{\psi_{+out}}{\psi_{+in}}\right\rvert }^2e^{-2i\pi (\delta-2)},  \\ \nonumber 
={\left\lvert \frac{\psi_{+out}}{\psi_{+in}}\right\rvert }^2e^{{4\pi}\left[2\tilde{\omega}+qQ\right]},
\end{align}
where we substitute $\delta$ as in \eqref{delta}.  
Using the fact that total probability of the particle wave going out from the horizon and the antiparticle wave going in must be equal to 1, we obtain this following distribution function, 
\begin{gather}
\left\langle \frac{\psi_{out}}{\psi_{in}}\mathrel{\left\lvert \vphantom{\frac{\psi_{out}}{\psi_{in}} \frac{\psi_{out}}{\psi_{in}}}\right.\kern-\nulldelimiterspace}\frac{\psi_{out}}{\psi_{in}}\right\rangle =1={\left\lvert \frac{A\phi_1}{B\phi_2}\right\rvert }^2\left\lvert 1-e^{{4\pi}\left[2\tilde{\omega}+qQ\right]}\right\rvert,  \label{exrad}
\end{gather}
and let us write the black hole's apparent horizon's radiation distribution function,
\begin{equation}
\left\lvert \frac{B}{A}\right\rvert^2=\zeta(T_H)=\frac{1}{e^{\frac{\hbar \omega}{k_B T_H}}-1}, \label{dist}
\end{equation}
where $k_B$ is the Boltzmann constant. The equation \eqref{exrad} in the near horizon limit, now can be rewritten as follows \cite{Sannan},
\begin{equation}
\zeta(T_H)=\left(\lim_{x\to 0} \frac{\phi_1}{\phi_2}\right)\times\left\lvert 1-e^{{4\pi}\left[2\tilde{\omega}+qQ\right]}\right\rvert= 0. \label{cond}
\end{equation}

By comparing \eqref{dist} and \eqref{cond}, we can conclude that the extremal Reissner-Nordstr\"om black black hole's horizon has a zero Hawking temperature, thus, does not radiate. This is in contrast with the non-extremal Reissner-Nordstr\"om black hole where the apparent horizon’s radiation distribution function is obtained as \cite{Senjaya2},
\begin{equation}
\zeta(T_H)=\left\lvert \frac{B}{A}\right\rvert ^2=\frac{1}{e^{\frac{4\pi}{\delta_r}\left[\frac{E}{\hbar c}r^2_+ \right]}-1}=\frac{1}{e^{\frac{\hbar \omega}{k_B T_H}}-1},
\end{equation}
leads to a non-zero Hawking temperature,
\begin{gather}
T_H=\frac{\delta_rc\hbar }{4\pi k_B r^2_+},
\end{gather} 
where $\delta_r=r_+ -r_-$ is as defined in equation \eqref{rpm}. One can also set $Q=M$ and directly substitute the expression into the Hawking temperature formula of the non-extremal Reissner-Nordstr\"om's above to get the same conclusion that the extremal Reissner-Nordstr\"om black black hole's horizon has a zero Hawking temperature.

\section{Summary}\label{sec:concl}
In this work, we explore the quasibound spectrum of charged relativistic bosonic quantum states of the massive and massless fields in an extremal Reissner-Nordstr\"om black hole space-time. We start with the Klein-Gordon equation with extremal Reissner-Nordstr\"om black hole background. Due to the static spherically symmetry, the temporal and the angular parts of the relativistic wave solution are obtained respectively in terms of the harmonic function and the Spherical Harmonics. We carefully deal with the radial equation and successfully find the exact solutions in terms of the double confluent Heun functions 

The quantization of the radial solution is done by applying the polynomial condition of the double confluent Heun function. The relation between the double confluent Heun's parameters turns out to be an energy quantization rule, presented in \eqref{energylevelsexact}. We also do a further investigation for the $\omega r_+\to 0$ limit and obtain the Hydrogenic atom's energy level \eqref{energylevelsapprox}. Similar expressions are also obtain via asymptotic methods, can be found in \cite{Pekeris,Sporea}. We also obtain exact energy level for  for massless charged scalar field around the extremal Reissner-Nordstr\"om black hole. We find that imaginary part of the quasibound state can either be positive, zero and negative while the real part can only be positive if $qQ<0$. Increasing scalar mass $\ti{\mu}$ leads to a more rapid decaying modes where the real part is increasing with $\ti{\mu}$. We can see that larger magnitude of $-qQ$ increases the combined attractive potential leads to enhancement of both the scalar fields' binding energy, $Re(\tilde{\omega})$ and the decay, which is represented by the magnitude of $-Im(\tilde{\omega})$.

Using the exact quantized energy expression \eqref{energylevelsexact}, we search for a possible charged scalar cloud around an extremal Reissner-Norstr\"om black hole. We found that there is no possible charged scalar cloud configuration for the extremal Reissner-Norstr\"om black hole. The same conclusion is also found in \cite{Garc}. We have also investigated the Hawking radiation of the extremal Reissner-Norstr\"om black hole's horizon via the Damour-Ruffini method. We find that the extremal static spherically symmetric charged black hole's horizon has zero temperature, thus, does not radiate. The same result is also obtained via black hole thermodynamics analysis in \cite{lust}.

\section*{Acknowledgments}
David Senjaya acknowledges this research project is supported by the Second Century Fund (C2F), Chulalongkorn University. SP acknowledges funding support from the NSRF via the Program Management Unit for Human Resources \& Institutional Development, Research and Innovation [grant number B39G670016].

\appendix
\section{The Normal Form and double confluent Heun differential equation} \label{AppendixA}
In this appendix, we discuss a useful transformation of a linear second order differential equation \cite{2420}. Let us first consider a general second order differential equation in the following form
\begin{align}
    \frac{d^2y}{dx^2}+\ti{p}(x)\frac{dy}{dx}+\ti{q}(x)y=0. \label{general ODE}
\end{align} 

Now, we assume that the solution can be written as
\begin{align}
        y=Y(x)e^{-\frac{1}{2}\int{\ti{p}(x)}dx}.
\end{align}

If one assumes that $e^{-\frac{1}{2}\int{\ti{p}(x)}dx}\neq 0$, then \eqref{general ODE} can be recast into the normal form 
\begin{align}
   \frac{d^2Y}{dx^2}+\left(-\frac{1}{2}\frac{d\ti{p}}{dx}-\frac{1}{4}\ti{p}^2+\ti{q}\right)Y=0.
\end{align}

Now, Let us consider the double confluent Heun differential equation \cite{Heun}, 
\begin{align}
\frac{d^2y}{dx^2}+\left(\frac{\gamma}{x^2}+\frac{\delta}{x}+\epsilon\right)\frac{dy}{dx}+\left(\frac{\alpha}{x}-\frac{\beta}{x^2}\right)y &=0. \label{DCHeq}
\end{align}

This equation has two irregular singular points at $x=0$ and $x=\infty$. A general solution of this equation is \cite{Heun}
\begin{align}
y &=A\operatorname{HeunD}\left(\beta,\alpha,\gamma ,\delta,\epsilon,x\right) + Be^{\frac{\gamma}{x}-\epsilon x}x^{2-\delta}\operatorname{HeunD}\left(-2+\beta+\delta,\alpha-2\epsilon,-\gamma ,4-\delta,-\epsilon,x\right)\nonumber\\
&\equiv \mathcal{H}(x),   
\end{align}
where $\operatorname{HeunD}$ is doubly confluent Heun function. One puts the double confluent Heun differential equation in its normal form by recognizing 
\begin{align}
\ti{p} &= \frac{\gamma}{x^2}+\frac{\delta}{x}+\epsilon,\\
\ti{q} &= \frac{\alpha}{x}-\frac{\beta}{x^2},\\
y &= Y(x)e^{-\frac{1}{2}\left(\epsilon x-\frac{\gamma}{x}\right)}x^{-\frac{\delta}{2}} \equiv \mathcal{H}(x),
\end{align}

Therefore, the normal form of \eqref{DCHeq} is obtained
\begin{align}
\frac{d^2Y}{dx^2}+K(x)Y &= 0,
\end{align}
where
\begin{align}
K(x) &= -\frac{1}{2}\frac{d\ti{p}}{dx}-\frac{1}{4}\ti{p}^2+\ti{q}, \nonumber \\
&= -\frac{\epsilon^2}{4}+\frac{1}{x}\left(-\frac{\epsilon \delta}{2}+\alpha\right) + \frac{1}{x^2}\left(\frac{\delta}{2}-\frac{\delta^2+2\epsilon \gamma}{4}-\beta\right) \nonumber \\
&~~~~~+\frac{1}{x^3}\left(\gamma-\frac{\delta \gamma}{2}\right)+\frac{1}{x^4}\left(-\frac{\gamma^2}{4}\right). \label{normalDCHeq}
\end{align}

Remark that the double confluent Heun function becomes polynomial when the following condition is met
\begin{align}
\frac{\alpha}{\epsilon} &= -n_r,\quad n_r=0, 1,2,\dots. \label{HeunDPol}   
\end{align}


\bibliography{sn-bibliography}

\end{document}